\documentclass{article}
\usepackage{spconf,amsmath,graphicx,bm,booktabs}
\usepackage{url,times,booktabs, tabularx}
\usepackage{graphicx}
\usepackage[colorlinks=false,pdfborder={0 0 0}]{hyperref}
\usepackage{units}

\usepackage{multirow,array}
\usepackage{balance}
\usepackage{soul}
\usepackage{color}
\usepackage{cleveref}
\usepackage{wrapfig}
\usepackage{booktabs}
\usepackage{balance}
\usepackage[activate]{microtype}
\usepackage{subfig}

\usepackage{wrapfig}
\usepackage{tabularx}
\usepackage{colortbl}
\usepackage{balance}
\usepackage{subfig}

\newcommand{\eg}{e.\,g., }
\newcommand{\ie}{i.\,e., }

\newcommand{\ds}{\textsc{Deep Spectrum\,}}

\newcommand{\asw}{\textsc{ASW\,}}

\usepackage{soul}
\definecolor{purp}{rgb}{0.3,0.1,0.4}
\definecolor{blau}{rgb}{0.4,0.4,0.9}

\ninept

\title{Acoustic Sounds for Wellbeing: A Novel Dataset and Baseline Results}

\name{Alice Baird$^{1}$,
      Bj\"orn Schuller$^{1,2}$
     }

\address{$^1$ ZD.B Chair of Embedded Intelligence for Health Care and Wellbeing, University of Augsburg, Germany \\
          $^2$ GLAM -- Group on Language, Audio \& Music, Imperial College London, UK\\
         {\small \tt alice.baird@informatik.uni-augsburg.de}
}
\begin{document}
%
\maketitle
\begin{abstract}
The field of sound healing includes ancient practices coming from a broad range of cultures. Across such practices there is a variety of acoustic instrumentation utilised. Practitioners suggest that sound has the ability to target both mental and even physical health issues, e.g., chronic-stress, or joint-pain. Instruments including the Tibetan singing bowl and vocal chanting, are still widely used today. With the noise-floor of modern urban soundscapes continually increasing and known to impact wellbeing, methods to improve this are needed. With that in mind, this study presents the Acoustic Sounds for Wellbeing (ASW) dataset. The ASW dataset is a dataset gathered from YouTube including 88\,+ hrs of audio from 5-classes of acoustic instrumentation (Gongs, Drumming, Singing Bowls, and Chanting). We additionally present initial baseline classification results on the dataset, finding that conventional Mel-Frequency Cepstra coefficient features achieve at best an unweighted average recalled of 57.4\% for a 5-class support vector machine classification paradigm.  
\end{abstract}
\section{Introduction}
\label{sec:intro}
The soundscape is an often overlooked aspect of our environment, and in urban scenarios the soundscape is becoming increasingly chaotic~\cite{atkinson2007ecology},leading to health issues for those in constant contact~\cite{kang2018soundscape}. Traffic, roadworks, construction, loud conversations, sirens, such a sonic cacophony results in conditions including insomnia, chronic stress, and hearing loss~\cite{brown2015role}. In this regard researchers are exploring methods for improving the audible environments, including through the integration of more \emph{pleasant} audio, at a level of \emph{sonic harmony}~\cite{jang2005selection}.

By definition, the soundscape is all the audio we hear at a given moment~\cite{Schafer1993}. The term was initially an anthropological concept, a tool for discussing historical cultural events, much like the landscape is for vision~\cite{Schafer1993}. The soundscape, can include both acoustic and synthetic-based audio, composed to fill an environment with a specific intention, \eg for environments in which excessive background noise is causing task-distraction~\cite{iyendo2016exploring}. 

Despite the impact of urban soundscapes, the healing ability of sound is well-known~\cite{gaynor1999sounds}. Acoustic instrumentation is prevalent in such scenarios, \eg the use of gongs, and drums, are common for altering states of consciousness, \eg the Tibetan singing bowls~\cite{conley2012holistic}, or the Native American ritualistic drumming~\cite{winkelman2003complementary}. With less attention as compared to acoustic instrumentation, synthetic audio is also applied for healing -- able to alter their current state of mind, \eg white noise inducing sleepiness~\cite{spencer1990white} and pure tones have shown to evoke arousal~\cite{Baird-18-PEI}. Synthetic audio, when combined with bio-acoustics, such as the heartbeat, have a significant affect on the listener, in particular encouraging \emph{individualism}~\cite{Parada-Cabaleiro17-SOP}. 

In regards to rhythm, it is sonic oscillations which synchronise with bodily function as humans are intrinsically rhythmical organismse~\cite{crowe1996overview}. In a similar way, repetitive methods including vocal stimulation, \eg singing, are able to maintain positive emotion~\cite{Olivier2015-CreateMusic}. As well as this instruments such as the gong which produce slow decaying oscillation have shown to have a calming affect, increasing mental awareness~\cite{Cook1995}. Wellbeing from music specifically, is also a prominent area of research~\cite{Raymound2013-MusicHealthAndWellbeing}. However much of the positivity evoked by music is known to come from stimulation of nostalgia, rather than the audio signals properties alone~\cite{barrett2010music}. For example, states of \emph{trance} are induced from repitious dance music, however it is known to be enhanced through the affect of social bonding~\cite{freeman1998neurobiological}. 

Augmenting a negative soundscape, through a combination of acoustic knowledge, and state-of-the-art computational approaches, could have great benefits in scenarios including busy offices, and hospitals~\cite{ahn2015analyses}. State-of-the-art approaches for audio generation are now not only showing great fidelity for both music and speech~\cite{Ping2018-DeepVoice}, but generative networks are now able to be conditioned, learning from multiple sources to tailor the generated audio more precisely~\cite{tobing2019voice}. For example emotional speech states can be adapted from one emotion to another, utilising a WaveNet Vocoder adapted framework~\cite{choi2019emotional}. As well as finding that an array of audio-based tasks including obtaining representations of emotional speech, are more effectively managed by deep convolutional generative adversarial networks (DC-GANs) as compared to convolutional neural networks (CNNs) architectures, when focusing on data (derived from audio) generation~\cite{chang2017learning}.


Such deep approaches require large quantities of data and to the best of the authors knowledge, there is currently no dataset available to the academic community in the realm of acoustic audio for wellbeing. With this in mind, this study presents the Acoustic Sounds for Wellbeing (\asw) dataset. \asw is a large dataset of 88+hrs of audio data, which has been gathered from YouTube. \asw is gathered as a means of exploring both the perception of audio, and as a tool for soundscape augmentation, and generation, and includes instrument classes known to promote states of positive wellbeing; such as Gongs, Drums, Chimes, Singing Bowls and Chanting. As a means of showing the variation between classes, and the efficacy of the \asw dataset for future research, we present a support vector machine (SVM) based classification experiment across varying class combinations. Utilising conventional Mel-Frequency Cepstra coefficents and state-of-the-art features we perform a series of multi-class classification tasks. 

This paper is organised as follows. The proceeding section  (\Cref{sec:dataset}), introduces the \asw dataset. We then describe our experimental settings for the baseline experiments in~\Cref{sec:exp}, including the proceeding classification paradigm. Following this we discuss the classification results in~\Cref{sec:res_duc} and finally, conclusions and future work plans are given in~\Cref{sec:conc}.

\section{The Acoustic Sounds for Wellbeing Dataset}
\label{sec:dataset}
The \asw dataset is a dataset collected from YouTube videos which have been licensed under creative commons. The full \asw dataset is available to the research community
~\footnote{link will be added upon acceptance} and has a total of 88\,:05\,:23 (hh\,:mm\,:ss,), from 68 audio clips (extracted from videos), across 5 classes (Gongs, Drumming, Chimes, Singing Bowls, Chanting. Based on our observation of the data, it can be seen that in general audio such as this (\ie sourced from YouTube), is of extensive duration (ca. 1hr per video), often changing slowly over time, and used for adapting the ambience. 

\vspace{-0.2cm}
\subsection{The ASW Collection Approach}
\label{sec:collection}
The audio data of \asw was extracted from Youtube\footnote{\url{https://youtube.com}} videos which are licensed under creative commons. Crawling was applied using the open source software youtube-dl\footnote{\url{https://ytdl-org.github.io/youtube-dl/index}}, applying the following search-terms, to download audio from the first 20 pages of each term: \textit{Gong+Sound+Healing}, \textit{Drums+Sound+Healing}, \textit{Chimes+Sound+Healing}, \textit{Singing+Bowls+Sound+Healing}.

\vspace{-0.2cm}
\subsection{Post Processing}
After the initial data download there were a series of processing steps which were applied to the data in order to make it more useable for further computational analysis. Firstly, files were renamed from there original video title, to a class label and instance value (\ie 01-chi.wav), with original video names stored for later reference purposes. Originally the audio was downloaded at a standard 44.1\,khz, 16 bit stereo WAV format, and for ease of processing this was converted to a mono WAV format for the final dataset. 

Due to the brute-force nature of the extraction process, as well as the use of lower-quality typically creative commons videos, there were many unusable videos for each of the classes within the dataset. As a means of cleaning the classes, the files were listened over briefly to check the quality and presence of any additional audio which may not fit the characteristics of the class \eg excessive speech. The human voice was a very common occurrence, due to aspects such as tutorials or live performances found via YouTube, and give this prevalence it has been difficult to remove all audio files with the presence of speech for this first iteration of the dataset. This process approximately halved the original dataset size, however with many audio files still of interest, this resulted in a new class, namely \emph{chanting} (cf. \Cref{sec:class} for a description of the classes). 

After this manual selection, and given the unmanageable duration of most files in the data set ($\mu$ 01\,:21\,:14), we chunked the data at segments of 60 seconds. The data was then normalised across all instances, and any files containing silence were removed automatically.

\begin{figure*}[htbp]
\centering
\subfloat[\tiny test/gon/g60c\_000\_55-gon]{\includegraphics[width=0.2\linewidth]{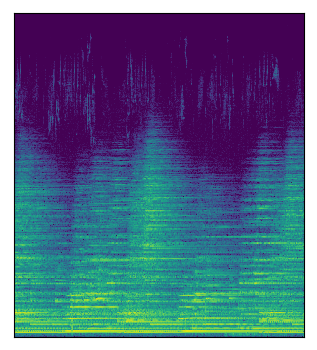}  }
\subfloat[\tiny test/dru/60c\_001\_64-dru]{\includegraphics[width=0.2\linewidth]{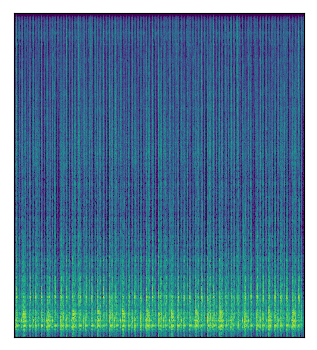} } 
\subfloat[\tiny train/chi/60c\_006\_38-chi]{\includegraphics[width=0.2\linewidth]{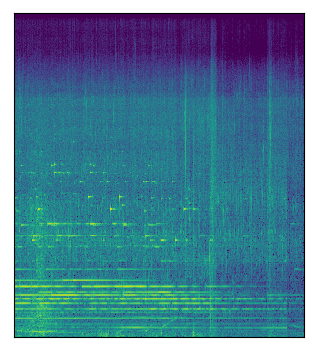} }
\subfloat[\tiny train/sin/60c\_003\_9-sin]{\includegraphics[width=0.2\linewidth]{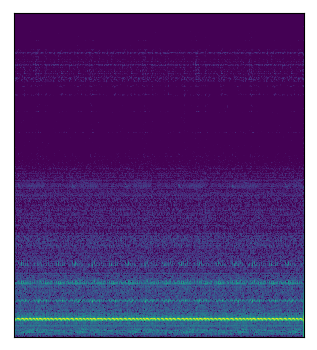}}
\subfloat[\tiny test/cha/60c\_002\_80-cha]{\includegraphics[width=0.2\linewidth]{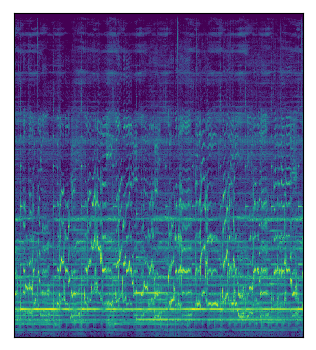}  }
\caption{Mel-spectrogram representations of the 60 second audio chunks for each audio class (a) Gong (b) Drumming (c) Chimes (d) Singing Bowl (e) Chanting.  Audio file names are indicated in the caption.}
    \label{fig:specimgs}
    \vspace{-0.3cm}
\end{figure*}

\vspace{-0.2cm}
\subsection{The Classes}
\label{sec:class}

Initially, when gathering the \asw dataset, classes were defined by the key terms which are mentioned in \Cref{sec:collection}. These classes were not only chosen for their prevalence in holistic healing but also their \emph{oscillatory} \ie repetitive nature, which has shown to be a fundamental aspect of healing sounds~\cite{crowe1996overview}. However during the processing phase, a reasonably large amount of \emph{chanting} was found, and so this was added as an additional class. We also found an abundance (ca. 20\,hrs) of weather sounds \eg rain, which have been discarded from this iteration of the dataset as they fall outside the current research scope. However, such audio would be of interest in future work, as healing from natural elements is also well known in the literature \cite{gillis2015review}. A visual representation of an audio instance for each class is shown in \Cref{fig:specimgs}, and a full description of the resulting class distribution is given in \Cref{tab:distribution}\footnote{Audio files depicted in \Cref{fig:specimgs} can be heard at the following link \url{http://tiny.cc/uohwez}.}. In this section we will discuss the positing of each class selection in the frame of wellbeing, and the type of audio which has been collected for this class.

\textbf{The Gong}: Meditation gongs are traditionally known in regions including Burma, Indonesia or China as temple bells~\cite{Cook1995}. Typically struck with a mallet, the low frequencies and slow oscillation decay, is known to results in a calming affect on the brain, whilst also increasing mental awareness, and promoting relaxation~\cite{Cook1995}. In the field of music therapy, it is known that the gong is able to stimulate imagery, free projection of feelings, and free aspects such as play and improvisation~\cite{Moreno-PSC}. The pattern of oscillation that is produced from the gong, has also begun to be researched for its affect on biological signalling and physical healing~\cite{muehsam2014life}.

\textbf{Drumming:} Drumming is known to improve focus, and alter mental states~\cite{kjellgren2010altered}. In this dataset, since we have the inclusion of \emph{sound+healing} we exclude electronic or standard drum set drumming, and appear to have mostly obtained audio from holistic settings, such as shaman rituals. During a shaman ritual, the drum is a single headed frame, sometimes with metallic objects hanging inside. The drum is known as a spirit, and drumming occurs over night throughout the ritual~\cite{hoskins1988drum}. Native American ceremonies, also utilise drumming, for states of consciousness alteration~\cite{thomason2010role}. 

\textbf{Wind-Chimes:} The wind-chime, is known to encourage deep meditative states due to its repetitious tones, which synchronise with natural weather patterns~\cite{gaynor1999sounds}, and is used commonly in calming outdoor scenarios~\cite{ahn2015analyses}. Often found in the foreground of more natural scenarios, it was this class which was dominated by additional weather sounds (excluding wind), \eg thunder, rain, and so where possible have excluded these instances in the first iteration of the dataset, as this may have lead to excessive variety in the class samples, making classification a challenge.

\textbf{The Singing Bowl:} The singing bowl has a few variations, but most common of the \emph{oriental} singing bowls, is the Tibetan singing bowl, typically made from seven \emph{holy} metals, resulting in a rich and vibrant oscillation when a mallet is continually played around the bowls rim.  Known as one of the most powerful healing instruments, and much like many other healing instruments, it is the repetitious nature of the playing which benefits both the player them self (encouraging gestural control) and those meditating in earshot\cite{crowe1996overview}. As is quite common with singing bowls, the player will play multiple tones, whilst leaving the last to resonate out~\cite{shrestha2009heal}, and therefore the audio gathered for this class, is of longer duration, and of sustained playing from one tone to another. 

\textbf{Vocal Chanting:}
Much like drumming, it is the repetitive nature of chanting which has healing properties~\cite{heather2007sound}. Although the heard chant is also beneficial when in a group setting \eg as a way of encouraging space sharing~\cite{bornholdt2010chanting}, it is generally the one chanting who is receiving a large affect from the experience, with research showing that regular practice of chanting can dramatically reduce stress~\cite{wolf2003examining}. Commonly, within this dataset there is chanting from shamanic ritual as a result of the drumming class keyword search. Therefore it is often not pure chanting. During shamin chanting the chants focus on the repetition of specific phrases, to encourage healing of an individual, \eg offering an infant child strength to heal~\cite{brown1988shamanism}.

\begin{table}[t!]
\centering
\footnotesize
\caption{Data independent partitions, train, (val)idation, test - class distribution of the \asw dataset post-processing; (gon)g, (dru)ms, (chi)mes, (sin)ging bowl, and (cha)nting. Including the number of 60 second segments (\# segs), and the total duration in hours:minutes(hh:mm).}
\label{tab:distribution}
\begin{tabular}{l|r|r|r|r}
\toprule
& Train & Val. & Test & $\sum$ \\
\midrule
\# videos & 24 & 22 & 22 & 68 \\
\midrule
gon & 12 & 11 & 11 & 34\\
dru & 13 & 11 & 11 & 35 \\
chi & 13 & 11 & 11 & 35 \\
sin & 12 & 11 & 11 & 34 \\
cha &5 & 5 & 5 & 15 \\
\midrule 
\# segs  & 1357 & 1971 & 1955 & 5283 \\
hh\,:mm  & 22\,:37 &32\,:51 & 32\,:35 &88\,:03\\
\bottomrule
\end{tabular}
\end{table}

\section{Experimental Setting}
\label{sec:exp}

As a first step for analysis of the classes within the \asw dataset we perform a series of baseline classification experiments: 

\begin{enumerate}
    \item \textbf{2 Class:} Gong (gon) vs. Singing Bowl (sb) - Given the similarity of these two instruments, such as the common use of materials (\ie often precious metal), we choose this combination to take a closer look at the difference in the data sources for these two classes.
    \item \textbf{3 Class:} Chimes (chi), Drum (dru), Gong - Counter to the 2 Class task, we choose this combination due to their intrinsic acoustic difference. 
    \item \textbf{4 Class:} Chimes, Drum, Gong, Singing Bowl - In this experiment we remove the Chanting class, as this is the only human related audio sample, as well as the nosiest (\ie most variety) and smallest class.   
    \item \textbf{5 Class:} Chanting (cha), Chime, Drum, Gong, Singing Bowl - All classes of the \asw dataset together, should be seen as the official baseline for the \asw dataset.  
\end{enumerate}

\begin{figure*}[htbp]
\centering
\subfloat[]{\includegraphics[width=0.25\linewidth]{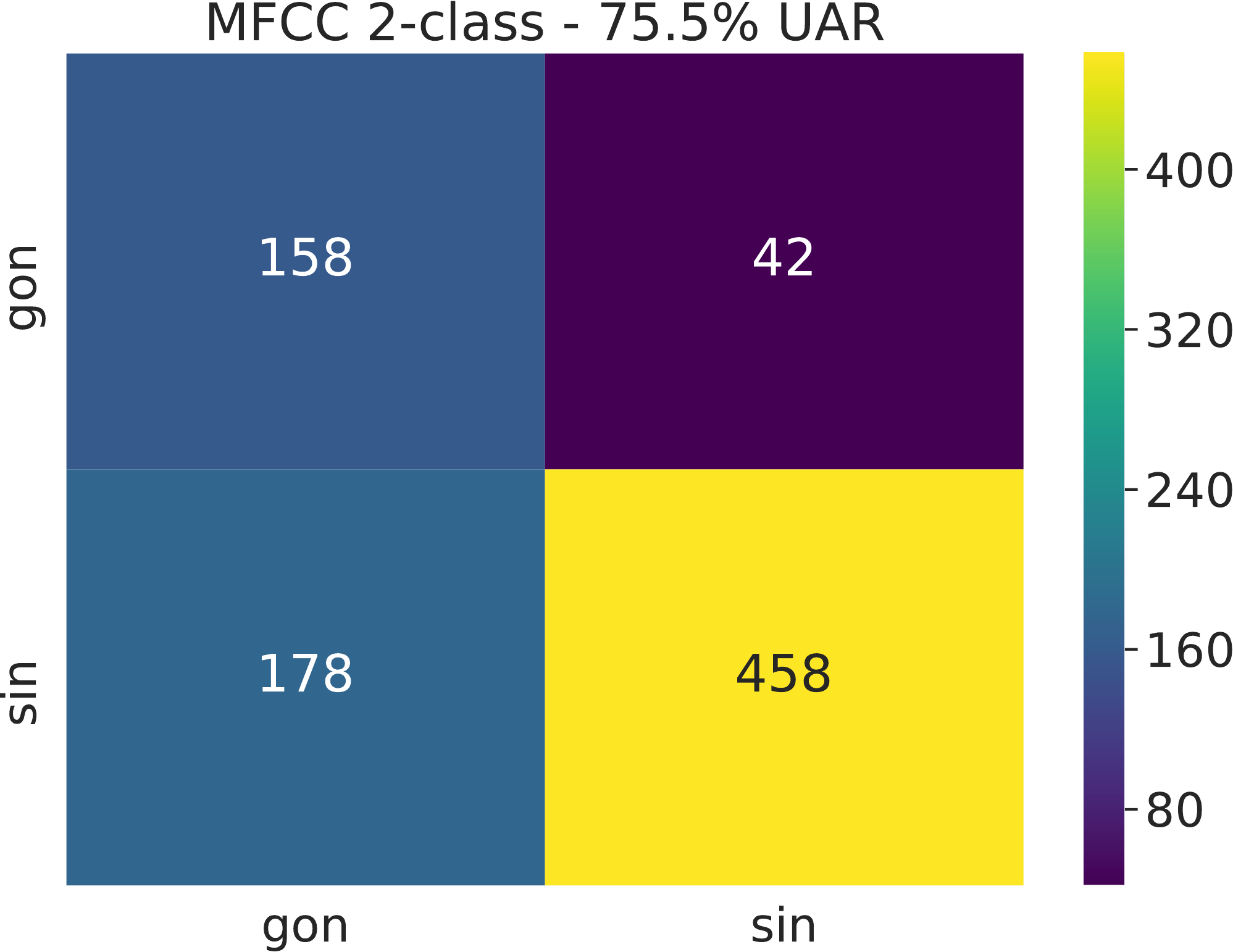}  }
\subfloat[]{\includegraphics[width=0.25\linewidth]{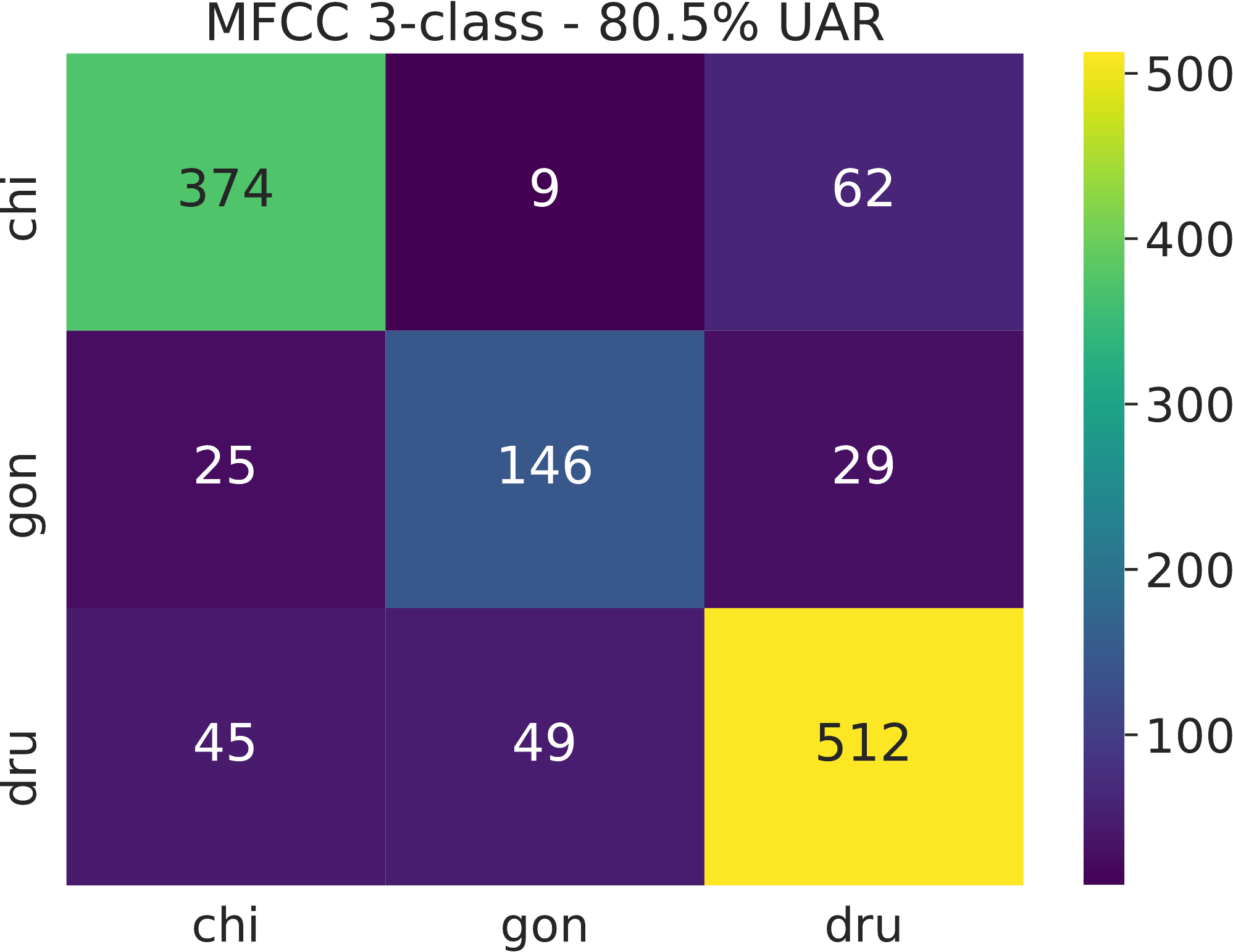} }
\subfloat[]{\includegraphics[width=0.25\linewidth]{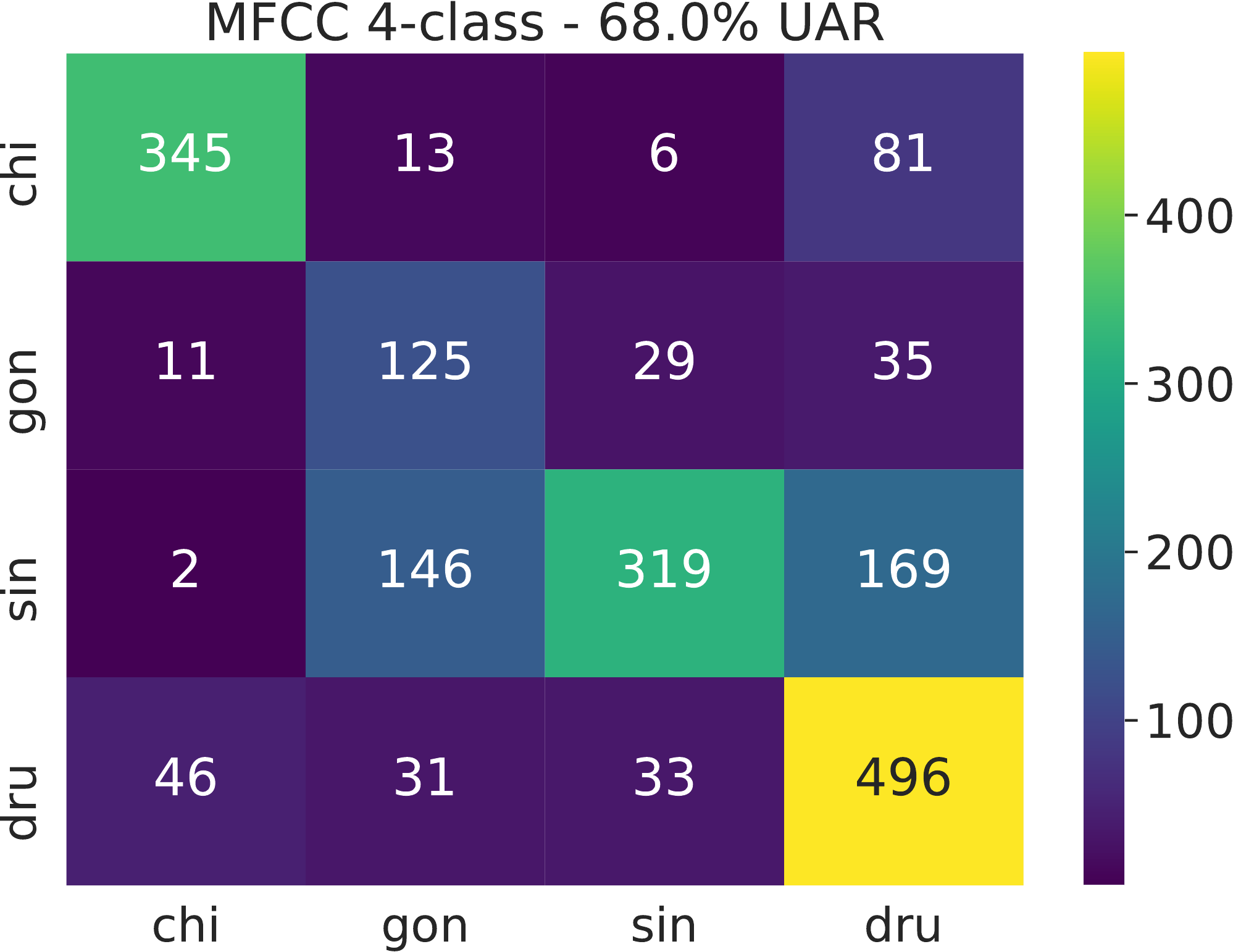} } 
\subfloat[]{\includegraphics[width=0.25\linewidth]{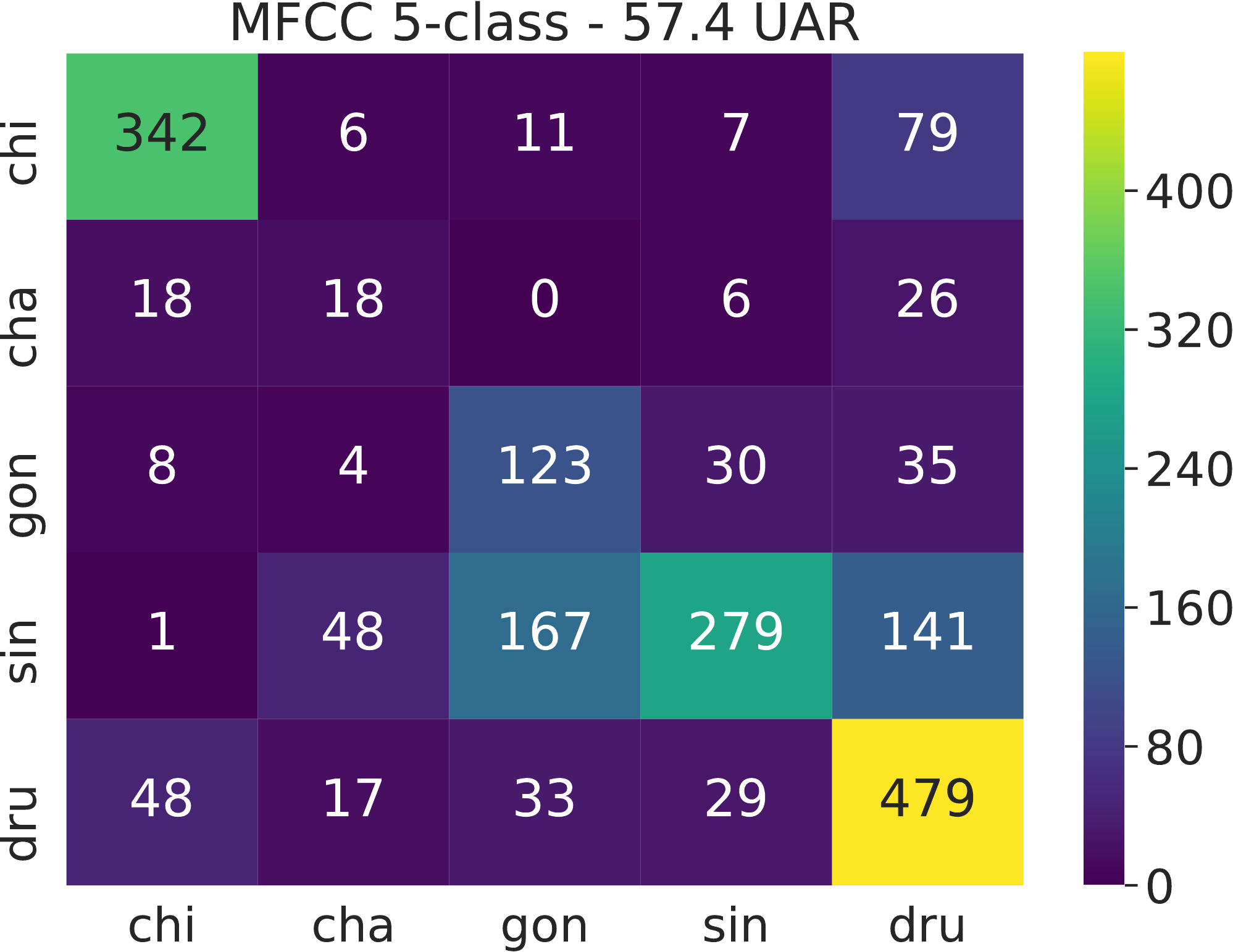}  }
   \caption{Confusion matrix, for each of the experiments 2,3,4 and 5-class classification paradigms, presenting the MFCC features set results only. Depicting that the limited data from class \emph{cha} impacts the overall classification results. }
    \label{fig:confusion}
    \vspace{-0.3cm}
\end{figure*}
\subsection{Baseline Partitioning}

To partition the data into train, validation and test, we chose a data-independent schema. Ultimately this led to a slight imbalance between classes due to the varied length of files. \Cref{tab:distribution} shows the distribution of instances over the 3 partitions, in total the \asw dataset contains (hh\,:mm), 88\,:03, which results in 22\,:37 for train, 32\,:51 for validation and 32\,:35 for the unseen testing set. In future iterations of the dataset it may be more fruitful to consider the total file size (or duration) for each partition rather than balancing video quantities.

\subsection{Feature sets} 
\label{sec:features}

For the experiments we extracted both conventional Mel-frequency cepstral coefficients (MFCC) features, as well as state of the art \ds features, in which deep data-representations from the audio signals are extracted.  

As a conventional approach, and since they have found great recognition across audio classifications tasks, including speech and music~\cite{guo2003content, weninger2011audio}, we utilise the LibROSA toolkit~\cite{mcfee2015librosa}, to extract all 20 MFCCs. MFCC features are able to capture a wide frequency range and incorporate the natural human hearing range which may be of benefit to this task given the acoustic nature of the instruments~\cite{weninger2011audio}. 

As a state-of-the-art audio approach, we extract 4\,096 deep data-representations using the \ds toolkit~\cite{amiriparian2017cnn}\footnote{\url{https://github.com/DeepSpectrum/DeepSpectrum}}. The \ds toolkit, extracts feature representations from the audio data using pre-trained convolutional neural networks. For this study we utilise the VGG19 pre-trained network. Extracting, viridis colour map, mel-spectrograms, the default \ds settings remain the same, and we extract over the full 60 seconds audio instance. 

\subsection{Classification Approach}
\label{sec:classify}
For our experiments, we use a support vector machine (SVM) with linear kernel implementation from the open-source machine learning toolkit Scikit-Learn~\cite{pedregosa2011scikit}. During the development phase, we trained a series of SVM models, optimising the complexity parameters ($C$ $\in$ $10^{-3}$, $10^{-2}$, $10^{-1}$, $1$), evaluating their performance on the validation set. Upon selection of the best performing complexity we report final test results on the unseen test partition against the concatenated train and validation sets.

\begin{table}[]
\centering
\footnotesize
\caption{Results for the 4 classification experiments (\#) as described in \Cref{sec:exp}, on the \asw dataset. Utilising a support vector machine (SVM), and optimising ($C$)omplexity, reporting unweighted average recall (\%) for both the Mel-frequency cepstral coefficients (MFCC) and \ds feature set (Dim)ensions. Providing also the chance level \% for each task. Emphasised results are discussed in \Cref{sec:res_duc}.}
\begin{tabular}{l|rrrrr}
\toprule
\multicolumn{6}{c}{SVM MFCC20}                                                               \\
\midrule
\#& Dim & Ch. & C & Dev. & Test \\
\midrule
(1)  & 20 &50 & $10^{-2}$ & 75.6  & 75.5\\
(2)  & 20 &33 & 1 & 34.8  &\textbf{80.5} \\
(3)  & 20 &25 &$10^{-1}$ &  28.1 &  68.0\\
(4)  & 20 &20 & $10^{-1}$ &32.7  &\textbf{57.4}  \\
\midrule
\multicolumn{6}{c}{SVM \ds VGG19} \\     
\midrule
\# & Dim &Ch & C. & Dev. & Test \\
\midrule
(1)  & 4\,096 &50\%& $10^{-1}$ & 63.5 &61.5  \\
(2)  & 4\,096 &33\%& $10^{-1}$ & 39.2  & \textbf{75.1}\\
(3)  & 4\,096 &25\%& $10^{-2}$ & 33.3 &\textbf{50.1}  \\
(4)  & 4\,096 &20\%& $10^{-2}$ & 34.0 & 47.3 \\
\bottomrule
\end{tabular}
\label{tab:results}
\end{table}

\section{Results an Discussion}
\label{sec:res_duc}
Results from the various experimental setups, are give in \Cref{tab:results}. We choose unweighted average recall (UAR) as our evaluation metric due to the imbalance between classes. 

Overall the MFCC results are stronger for each classification scenario on the test set. Although it is negligible, we do see improvement from \ds on the validation set. As a comparison between the features sets the standard MFCC results have shown to be robust against the state-of-the-art approach here, however further optimisation, including a smaller window size may prove more effective for the \ds features. 

In all classification scenarios we see a strong result above chance level, however when looking at the confusion matrix (cf. \Cref{fig:confusion}) we see that for the 5-class classification (experiment 4), the \textit{chanting} (cha) class is completely mis-classified, we assume this is due to the imbalance of the dataset. However it would be interesting to explore the acoustic nature of the chanting files given that they are unlike other classes, and the only human-based class. 
When looking at the 4-class results in the confusion matrix (cf. \Cref{fig:confusion})), we see that results between classes are strong, and in particular the \textit{drumming} (dru) class is highly different from all other classes. From \Cref{fig:specimgs} we can see that is most likely due to the drum strikes, which occur in the audio consistently representing a more unique acoustic space. 

\section{Conclusion and Future Work}
\label{sec:conc}
In this study, we have presented the \asw dataset, of 88\,+ hours of acoustic sounds which are known to encourage wellbeing gathered from YouTube. Results from a baseline classification experiement, have shown that at best a 5-class classification with MFCC features can achieve up to 57.4\% UAR which is 37 percent points above chance level. Although still achieving above chance level the \ds features do not achieve as strong results, suggesting the need for further and more fine grained data manipulation, and the large 60 second chunks may not have affective for this feature set.  

Given the large size of this data set, future work will include automatic generation of positive and/or calming soundscapes, as well as a deep human evaluation of the emotionality portrayed by each class, although given the quantity of data, this would require computational machine learning paradigms such as active learning. In this way working toward a method for conditional audio generation personalised to specific wellbeing needs. Motivated by the need to improve poorly designed urban environments, through further processing this data set will be of use for learning specific acoustic attributes which could be characterised during audio generation.

\section{Acknowledgements}
This work is funded by the Bavarian State Ministry of Education, Science and the Arts in the framework of the Centre Digitisation.Bavaria (ZD.B).

\footnotesize
\balance
\bibliographystyle{IEEEbib}
\bibliography{refs}

\end{document}